\documentstyle[12pt]{article}

\begin{document}
\vskip 72pt
\centerline{\large {\bf{IN MEDIUM EFFECTS ON THE $\phi$ MESON}}}
\vskip 36pt
\centerline{\bf Abhijit Bhattacharyya$^a$\footnote{Electronic Mail : 
phys@boseinst.ernet.in}, Sanjay K. Ghosh$^b$\footnote{Electronic Mail : 
sanjay@iopb.ernet.in}, S.C.Phatak$^b$\footnote{Electronic Mail : 
phatak@iopb.ernet.in} 
and Sibaji Raha$^a$\footnote{Electronic Mail : sibaji@boseinst.ernet.in}} 
\vskip 1pt
\centerline{a) Department of Physics, Bose Institute}
\centerline{93/1 A.P.C.Road}
\centerline{Calcutta 700 009, India}
\centerline{b) Institute of Physics}
\centerline{Sachivalaya Marg, Bhubaneswar - 751 005, India}
\vskip 36pt
\noindent \centerline {\bf Abstract}
\vskip 15 pt
The temperature dependence of the $\phi$ meson mass and its decay width 
($\phi \rightarrow K \bar{K}$)have been studied  from an effective
non-linear chiral Lagrangian in $SU(3)$.  Effective mass has been obtained
from the pole position of the full propagator. The width has also been
calculated. It has been found that the mass decreases with temperature
very slowly whereas the decay width increases quite sharply. Possible
consequence on the QGP signals is discussed. 
\newpage
\vskip 15 pt
The expected formation of a new  phase  of strongly interacting matter called 
Quark Gluon Plasma (QGP)\cite{a}, the restoration of the spontaneously 
broken chiral symmetry \cite {b} and other such tantalizing possibilities 
make the study of hot and dense hadronic matter a vibrant area of research. 
QGP, if it is formed, exists only for a fraction of the total evolution time. 
So to look for the signals of QGP is both an interesting as well as 
challenging field of research. This is particularly so because one has to 
disentangle the signals coming form the hadronic sector. For example, it is 
believed that dileptons and photons are good signals of QGP \cite{c}. The 
interactions of these particles with the surroundings are electromagnetic in 
nature. Hence, they are affected the least by the final state interactions, 
enabling them to bring out information from the core of the plasma. However 
these particles are also produced in hadronic processes and one has to 
subtract the hadronic contribution to get the information about QGP. 
Therefore the study of effective meson masses \cite{d} and decay widths at
finite  
temperature is of high current interest; in addition to having bearing on
the possible signatures of the 
putative phase transition, they may also provide important 
information about the state of the hot and dense hadronic matter formed in
ultrarelativistic heavy ion collisions.  

From the hadronic sector dileptons may be produced by reactions like $\pi^{+} 
\pi^{-} \rightarrow
\rho \rightarrow \mu^{+} \mu^{-}$ and/or 
$K \bar{K} \rightarrow \phi \rightarrow \mu^{+} \mu^{-}$. Other channels 
(like decays and binary reactions of various hadrons) may also be important, 
as argued by Gale and Lichard \cite{e}. They are however important for 
$M>1.5 GeV$. The $\pi^{+}\pi^{-}$ 
annihilation is very well studied at finite temperature. The study of 
$K \bar{K}$  
annihilation is interesting in the sense that the mass of $\phi$ meson 
is very close to twice the $K$ meson mass. Hence, even a small 
change in the $\phi$ meson or $K$ meson mass may have a strong effect 
on this process. If the $\phi$ mass falls below twice the $K$ mass 
then no $\phi$ peak will be observed in the dilepton spectra. 
Asakawa and Ko \cite{g} and Ko and Siebert \cite{h} have used this possibility 
to propose a new signal for the measurement of the transition temperature, 
namely the occurrence of a secondary $\phi$ peak. In this letter, we examine 
critically the validity of these premises.

Since hadrons cannot yet be  described  by QCD, due to our limited 
knowledge of its non-perturbative features,  we have to depend  on 
models. Hence, all the calculations so far are model dependent and the
results vary widely from model to model \cite{d}. Ko and collaborators 
\cite{g,h} based their conclusions on the temperature dependence 
of the $\phi$ mass calculated from the QCD sum rule. Here we will 
use the nonlinear ${\sigma}$ (NLS) model \cite{i,j}, which reproduces the low 
energy structure of QCD quite successfully. This is a purely hadronic model 
where confinement does not play any role. In our opinion, this provides a great
advantage over the Nambu-Jona-Lasinio model or other effective quark models.
In a previous work, the temperature dependence of $K$ mass has been studied 
\cite{k} in NLS with broken $SU(3)_V$. Here we will look at the temperature 
dependence of the $\phi$ meson mass and its decay width within the same level 
of approximations as in \cite{k}. Since we are primarily interested in 
studying the $\phi \rightarrow K\bar{K}$, the importance of $SU(3)$ in this 
context cannot obviously be stressed too strongly.

There are two popular and equivalent approaches to the chiral effective model; 
the Hidden Gauge Symmetry Approach (HGSA) \cite {j} and the Massive Yang Mills 
Approach (MYMA) \cite{i}. In this paper we will follow the HGSA. In this 
approach the vector mesons are incorporated as the dynamical gauge bosons of 
the hidden local symmetry contrary to the MYMA where the masses of the vector 
mesons are put in by hand. The HGSA Lagrangian, at the lowest order, 
can be written as a linear combination ${{\cal {L}}_A} + a {{\cal {L}}_V}$, 
$a$ being an arbitary parameter which is fixed from the condition of 
reproducing the Vector Meson Dominance (VMD). This condition leads to $a=2$. 
The Lagrangian in presence of $SU(3)_V$ breaking term is given by \cite {k},

\begin{eqnarray}
{{\cal {L}}_A} + \Delta {{\cal {L}}_A}   = -\frac {1}{8} f_{\pi}^2 Tr.[ 
{(D_{\mu}{{\xi}_L}.{{{\xi}_L}^{\dagger}}  - D_{\mu}{{\xi}_R}.{{{\xi}_R}^
{\dagger}})}^2 [ 1+ ({{\xi}_L}{{\epsilon}_A}{{{\xi}_R}^{\dagger}}
+{{\xi}_R}{{\epsilon}_A}{{{\xi}_L}^{\dagger}})]] \nonumber\\
{{\cal {L}}_V} +  \Delta{{\cal {L}}_V} = -\frac {1}{8} f_{\pi}^2 Tr. 
[{(D_{\mu}{{\xi}_L}.{{{\xi}_L}^{\dagger}}  
+ D_{\mu}{{\xi}_R}.{{{\xi}_R}^{\dagger}})}^2 [ 1+ ({{\xi}_L}{{\epsilon}_V}
{{{\xi}_R}^{\dagger}}+{{\xi}_R}{{\epsilon}_V}{{{\xi}_L}^{\dagger}})]]
\end{eqnarray}
The explanation of different terms can be obtained in ref. \cite{k}.
The QCD anomaly has been incorporated via the Wess-Zumino term. The 
parameters have been fitted with the experimental values of the $\phi$ mass
and the $\phi \rightarrow K\bar{K}$ decay width at zero temperature.

In the effective Lagrangian approach at zero temperature, it is  
assumed  that the properties of the system are describable at the tree level,
where the masses and the coupling constants are to be regarded  as  the  
physical ones. Loop diagrams, which are neglected here, produce only  
renormalization effects on them.

The calculation of effective mass follows the usual principle; Dyson's 
equation relates the free and the full propagators as 
\begin{equation}
D(p)  =  D_0 (p) + D_0 (p){\prod}D(p) 
\end{equation} 
where $D_0$  is the free propagator and $D$ is the full propagator. The 
effect of interaction is embedded in the polarisation function $\prod$. The 
temperature dependent polarisation has a real and an imaginary part. The real 
part contributes to the mass of the corresponding meson, while the imaginary 
part determines the decay width. Hence a  self consistent solution  of the 
equation 
\begin{equation}
\omega^2  - m_{\phi}^2 - Re[\prod_{\phi} (\omega,|\bf k|\longrightarrow 0)]
{\big |_{{k_0}={m_{\phi}^*}}} = 0
\end{equation}
gives the temperature dependence of $\phi$ meson mass. In eq.(3) $\omega = 
{\sqrt {{|\bf k|}^2 + {m_{\phi}^{*2}}}}$ {\hskip 0.05in} is the 
effective energy and ${m_{\phi}^*}$ is the effective mass of the $\phi$
meson. We have calculated the self energy in the one loop approximation with 
$K \bar{K}$ and $\rho \pi$ loop. Diagrams with heavy vector mesons in both 
the internal lines have been neglected as they will be Boltzman suppressed at 
finite temperature. For the sake of brevity, we have not shown the calculation 
of the self energy; these details may be obtained from ref.\cite{d}. 

The  $\phi$ pole  mass  as  a function of temperature has been plotted in 
Fig.~[1]. The thermal decay width ($\phi \rightarrow K \bar{K}$) has been 
plotted in Fig.~[2]. It can be readily seen from Fig.~[1] that the mass 
decreases very slowly with temperature. This result has an excellent agreement 
with the experimental data from the AGS \cite{l} which have become available 
rather recently. For the most central collisions ($T = 150 MeV$), the 
$\phi$-meson mass is found to decrease to about $1016 MeV - 1017 MeV$ 
(depending on where the $p_T$ cut has been applied). At that temperature our  
result gives a value of $1016 MeV$. In ref. \cite{g}, the temperature 
dependence of $\phi$ mass has been calculated using the QCD sum rule. It has 
been reported there that the mass decreases with temperature and goes below 
twice $K$ meson mass above $T=150MeV$ which will make the $\phi \rightarrow 
K \bar{K}$ process forbidden \cite{g}. But, the validity of QCD sum rule, 
which is based on Operator Product Expansion (OPE), is questionable at such 
a high temperature. According to ref.\cite{m}, QCD sum rule calculation is 
valid only upto $100 MeV$ {\it i.e.}, when the parameter 
$\epsilon = T^2/6{f^2_\pi}$ is small. Furthermore, the QCD sum rule is not a 
pure hadronic picture and the relation with the hadrons is built in through 
the {\it ansatz}  of parton-hadron duality. Moreover, these results 
fail to reproduce the experimental conclusion, even qualitatively, that the 
change in $\phi$-meson mass should be small due to in medium effects \cite{l}. 
Even more importantly, Ko and collaborators ignored the temperature 
dependence of the $K$ mass entirely. In our calculation
we find that $\phi$ meson mass always remains above the two kaon
threshold; as a result, the effect of $\phi \rightarrow K\bar{K}$ decay
width will be important. 

As seen from Fig.~[2], the decay width of the $\phi$-meson increases with 
temperature quite sharply. In fact this result has a strong dependence on the 
temperature dependence of the $\phi$ and $K$-meson masses. The decay width 
increases from $3.7 MeV$ ( at zero temperature) to about $21 MeV$. In ref. 
\cite{h}, the decay width becomes zero at about temperature $150 MeV$. Our
calculations show that this is entirely due to the neglect of the
temperature dependence of kaon mass. The small decrease of kaon mass leads to 
an increase in the decay width of $\phi$ meson, even if the $\phi$ mass 
decreases. This is the first time that the $\phi$ and $K$ meson masses have 
been calculated self consistently from a full $SU(3)$ chiral effective 
Lagrangian. Our result shows that this self consistency is strongly required 
to determine the in-medium effects on the $\phi$ meson. 

The above results have a strong bearing on the QGP diagnostics. One must 
immediately conclude that a primary $\phi$-peak can indeed be observed in the 
dilepton spectrum from the hadronic sector. Obviously, the secondary peak
expected from the collisional broadening of the $\phi$-width \cite{h} would
be overshadowed by the primary peak. The mass shift as found in the current
data is in excellent agreement with the theoretical prediction of our model.

The work of AB has been supported in part by the Department of Atomic Energy 
(Government of India). 
\newpage

{\bf{FIGURE CAPTIONS:}}

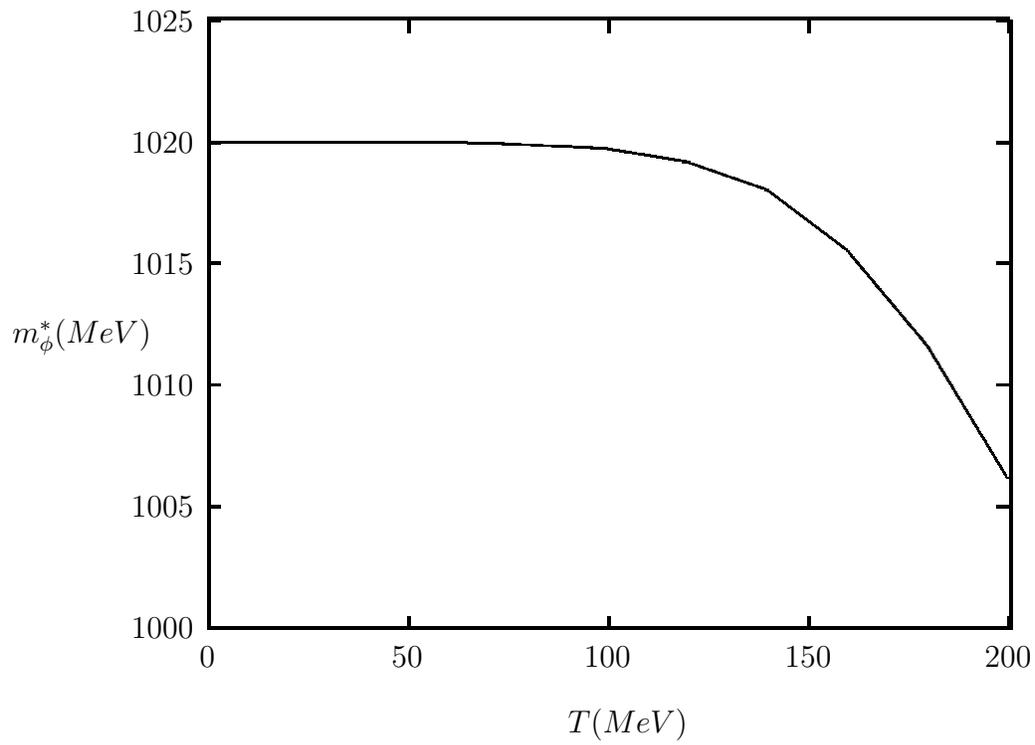
Figure 1: Temperature dependence of the $\phi$ mass.

Figure 2: Temperature dependence of the $\phi$  decay width.

\newpage
\begin{figure}
\setlength{\unitlength}{0.240900pt}
\ifx\plotpoint\undefined\newsavebox{\plotpoint}\fi
\sbox{\plotpoint}{\rule[-0.500pt]{1.000pt}{1.000pt}}%
\begin{picture}(1500,1049)(0,0)
\font\gnuplot=cmr10 at 10pt
\gnuplot
\sbox{\plotpoint}{\rule[-0.500pt]{1.000pt}{1.000pt}}%
\put(176.0,68.0){\rule[-0.500pt]{1.000pt}{230.782pt}}
\put(176.0,68.0){\rule[-0.500pt]{4.818pt}{1.000pt}}
\put(154,68){\makebox(0,0)[r]{$1000$}}
\put(1416.0,68.0){\rule[-0.500pt]{4.818pt}{1.000pt}}
\put(176.0,259.0){\rule[-0.500pt]{4.818pt}{1.000pt}}
\put(154,259){\makebox(0,0)[r]{$1005$}}
\put(1416.0,259.0){\rule[-0.500pt]{4.818pt}{1.000pt}}
\put(176.0,450.0){\rule[-0.500pt]{4.818pt}{1.000pt}}
\put(154,450){\makebox(0,0)[r]{$1010$}}
\put(1416.0,450.0){\rule[-0.500pt]{4.818pt}{1.000pt}}
\put(176.0,641.0){\rule[-0.500pt]{4.818pt}{1.000pt}}
\put(154,641){\makebox(0,0)[r]{$1015$}}
\put(1416.0,641.0){\rule[-0.500pt]{4.818pt}{1.000pt}}
\put(176.0,831.0){\rule[-0.500pt]{4.818pt}{1.000pt}}
\put(154,831){\makebox(0,0)[r]{$1020$}}
\put(1416.0,831.0){\rule[-0.500pt]{4.818pt}{1.000pt}}
\put(176.0,1022.0){\rule[-0.500pt]{4.818pt}{1.000pt}}
\put(154,1022){\makebox(0,0)[r]{$1025$}}
\put(1416.0,1022.0){\rule[-0.500pt]{4.818pt}{1.000pt}}
\put(176.0,68.0){\rule[-0.500pt]{1.000pt}{4.818pt}}
\put(176,23){\makebox(0,0){$0$}}
\put(176.0,1006.0){\rule[-0.500pt]{1.000pt}{4.818pt}}
\put(490.0,68.0){\rule[-0.500pt]{1.000pt}{4.818pt}}
\put(490,23){\makebox(0,0){$50$}}
\put(490.0,1006.0){\rule[-0.500pt]{1.000pt}{4.818pt}}
\put(804.0,68.0){\rule[-0.500pt]{1.000pt}{4.818pt}}
\put(804,23){\makebox(0,0){$100$}}
\put(804.0,1006.0){\rule[-0.500pt]{1.000pt}{4.818pt}}
\put(1119.0,68.0){\rule[-0.500pt]{1.000pt}{4.818pt}}
\put(1119,23){\makebox(0,0){$150$}}
\put(1119.0,1006.0){\rule[-0.500pt]{1.000pt}{4.818pt}}
\put(1433.0,68.0){\rule[-0.500pt]{1.000pt}{4.818pt}}
\put(1433,23){\makebox(0,0){$200$}}
\put(1433.0,1006.0){\rule[-0.500pt]{1.000pt}{4.818pt}}
\put(176.0,68.0){\rule[-0.500pt]{303.534pt}{1.000pt}}
\put(1436.0,68.0){\rule[-0.500pt]{1.000pt}{230.782pt}}
\put(176.0,1026.0){\rule[-0.500pt]{303.534pt}{1.000pt}}
\put(930,-198){\makebox(0,0)[r]{\bf {Figure 1}}}
\put(930,-84){\makebox(0,0)[r]{$T(MeV)$}}
\put(88,526){\makebox(0,0)[r]{$m_\phi^* (MeV)$}}
\put(176.0,68.0){\rule[-0.500pt]{1.000pt}{230.782pt}}
\sbox{\plotpoint}{\rule[-0.300pt]{0.600pt}{0.600pt}}%
\put(176,831){\usebox{\plotpoint}}
\put(553,829.25){\rule{15.177pt}{0.600pt}}
\multiput(553.00,829.75)(31.500,-1.000){2}{\rule{7.588pt}{0.600pt}}
\multiput(616.00,828.50)(11.279,-0.501){13}{\rule{12.683pt}{0.121pt}}
\multiput(616.00,828.75)(161.675,-9.000){2}{\rule{6.342pt}{0.600pt}}
\multiput(804.00,819.50)(3.068,-0.500){37}{\rule{3.750pt}{0.121pt}}
\multiput(804.00,819.75)(118.217,-21.000){2}{\rule{1.875pt}{0.600pt}}
\multiput(930.00,798.50)(1.442,-0.500){83}{\rule{1.868pt}{0.120pt}}
\multiput(930.00,798.75)(122.122,-44.000){2}{\rule{0.934pt}{0.600pt}}
\multiput(1056.00,754.50)(0.658,-0.500){185}{\rule{0.939pt}{0.120pt}}
\multiput(1056.00,754.75)(123.050,-95.000){2}{\rule{0.470pt}{0.600pt}}
\multiput(1182.00,657.41)(0.500,-0.595){247}{\rule{0.120pt}{0.864pt}}
\multiput(1179.75,659.21)(126.000,-148.206){2}{\rule{0.600pt}{0.432pt}}
\multiput(1308.00,506.27)(0.500,-0.826){247}{\rule{0.120pt}{1.140pt}}
\multiput(1305.75,508.63)(126.000,-205.633){2}{\rule{0.600pt}{0.570pt}}
\put(176.0,831.0){\rule[-0.300pt]{90.819pt}{0.600pt}}
\end{picture}
\end{figure}
\newpage
\begin{figure}
\setlength{\unitlength}{0.240900pt}
\ifx\plotpoint\undefined\newsavebox{\plotpoint}\fi
\sbox{\plotpoint}{\rule[-0.500pt]{1.000pt}{1.000pt}}%
\begin{picture}(1500,1049)(0,0)
\font\gnuplot=cmr10 at 10pt
\gnuplot
\sbox{\plotpoint}{\rule[-0.500pt]{1.000pt}{1.000pt}}%
\put(176.0,68.0){\rule[-0.500pt]{303.534pt}{1.000pt}}
\put(176.0,68.0){\rule[-0.500pt]{1.000pt}{230.782pt}}
\put(176.0,68.0){\rule[-0.500pt]{4.818pt}{1.000pt}}
\put(154,68){\makebox(0,0)[r]{$0$}}
\put(1416.0,68.0){\rule[-0.500pt]{4.818pt}{1.000pt}}
\put(176.0,256.0){\rule[-0.500pt]{4.818pt}{1.000pt}}
\put(154,256){\makebox(0,0)[r]{$5$}}
\put(1416.0,256.0){\rule[-0.500pt]{4.818pt}{1.000pt}}
\put(176.0,444.0){\rule[-0.500pt]{4.818pt}{1.000pt}}
\put(154,444){\makebox(0,0)[r]{$10$}}
\put(1416.0,444.0){\rule[-0.500pt]{4.818pt}{1.000pt}}
\put(176.0,632.0){\rule[-0.500pt]{4.818pt}{1.000pt}}
\put(154,632){\makebox(0,0)[r]{$15$}}
\put(1416.0,632.0){\rule[-0.500pt]{4.818pt}{1.000pt}}
\put(176.0,819.0){\rule[-0.500pt]{4.818pt}{1.000pt}}
\put(154,819){\makebox(0,0)[r]{$20$}}
\put(1416.0,819.0){\rule[-0.500pt]{4.818pt}{1.000pt}}
\put(176.0,1007.0){\rule[-0.500pt]{4.818pt}{1.000pt}}
\put(154,1007){\makebox(0,0)[r]{$25$}}
\put(1416.0,1007.0){\rule[-0.500pt]{4.818pt}{1.000pt}}
\put(176.0,68.0){\rule[-0.500pt]{1.000pt}{4.818pt}}
\put(176,23){\makebox(0,0){$0$}}
\put(176.0,1006.0){\rule[-0.500pt]{1.000pt}{4.818pt}}
\put(490.0,68.0){\rule[-0.500pt]{1.000pt}{4.818pt}}
\put(490,23){\makebox(0,0){$50$}}
\put(490.0,1006.0){\rule[-0.500pt]{1.000pt}{4.818pt}}
\put(804.0,68.0){\rule[-0.500pt]{1.000pt}{4.818pt}}
\put(804,23){\makebox(0,0){$100$}}
\put(804.0,1006.0){\rule[-0.500pt]{1.000pt}{4.818pt}}
\put(1119.0,68.0){\rule[-0.500pt]{1.000pt}{4.818pt}}
\put(1119,23){\makebox(0,0){$150$}}
\put(1119.0,1006.0){\rule[-0.500pt]{1.000pt}{4.818pt}}
\put(1433.0,68.0){\rule[-0.500pt]{1.000pt}{4.818pt}}
\put(1433,23){\makebox(0,0){$200$}}
\put(1433.0,1006.0){\rule[-0.500pt]{1.000pt}{4.818pt}}
\put(176.0,68.0){\rule[-0.500pt]{303.534pt}{1.000pt}}
\put(1436.0,68.0){\rule[-0.500pt]{1.000pt}{230.782pt}}
\put(176.0,1026.0){\rule[-0.500pt]{303.534pt}{1.000pt}}
\put(930,-307){\makebox(0,0)[r]{\bf {Figure 2}}}
\put(930,-119){\makebox(0,0)[r]{$T(MeV)$}}
\put(88,444){\makebox(0,0)[r]{$\Gamma (MeV)$}}
\put(176.0,68.0){\rule[-0.500pt]{1.000pt}{230.782pt}}
\sbox{\plotpoint}{\rule[-0.300pt]{0.600pt}{0.600pt}}%
\put(176,207){\usebox{\plotpoint}}
\put(490,206.25){\rule{15.177pt}{0.600pt}}
\multiput(490.00,205.75)(31.500,1.000){2}{\rule{7.588pt}{0.600pt}}
\put(553,207.25){\rule{15.177pt}{0.600pt}}
\multiput(553.00,206.75)(31.500,1.000){2}{\rule{7.588pt}{0.600pt}}
\multiput(616.00,209.99)(3.745,0.501){13}{\rule{4.350pt}{0.121pt}}
\multiput(616.00,207.75)(53.971,9.000){2}{\rule{2.175pt}{0.600pt}}
\multiput(679.00,219.00)(1.691,0.500){33}{\rule{2.139pt}{0.121pt}}
\multiput(679.00,216.75)(58.559,19.000){2}{\rule{1.070pt}{0.600pt}}
\multiput(742.00,238.00)(1.430,0.500){39}{\rule{1.841pt}{0.121pt}}
\multiput(742.00,235.75)(58.179,22.000){2}{\rule{0.920pt}{0.600pt}}
\multiput(804.00,260.00)(0.618,0.500){199}{\rule{0.891pt}{0.120pt}}
\multiput(804.00,257.75)(124.150,102.000){2}{\rule{0.446pt}{0.600pt}}
\multiput(930.00,362.00)(0.663,0.500){185}{\rule{0.946pt}{0.120pt}}
\multiput(930.00,359.75)(124.037,95.000){2}{\rule{0.473pt}{0.600pt}}
\multiput(1057.00,456.00)(0.500,0.600){245}{\rule{0.120pt}{0.870pt}}
\multiput(1054.75,456.00)(125.000,148.194){2}{\rule{0.600pt}{0.435pt}}
\multiput(1182.00,606.00)(0.500,0.571){247}{\rule{0.120pt}{0.836pt}}
\multiput(1179.75,606.00)(126.000,142.265){2}{\rule{0.600pt}{0.418pt}}
\multiput(1308.00,750.00)(0.500,0.695){247}{\rule{0.120pt}{0.983pt}}
\multiput(1305.75,750.00)(126.000,172.959){2}{\rule{0.600pt}{0.492pt}}
\put(176.0,207.0){\rule[-0.300pt]{75.643pt}{0.600pt}}
\end{picture}
\end{figure}

\newpage
\begin{thebibliography}{99} 
\bibitem{a} E.V.Shuryak, {\it Phys. Rep.} {\bf 61}, 71 (1980).
\bibitem{b} R.D.Pisarski, {\it Phys. Lett.} {\bf B110}, 155 (1982).
\bibitem{c} J.Alam, S.Raha and B.Sinha, {\it Physics Reports} (in press).    
\bibitem{d} A.Bhattacharyya and S.Raha, {\it J. Phys.} {\bf G21}, 741 (1995). 
\bibitem{e} C.Gale and P.Lichard, {\it Phys.Rev.} {\bf D49}, 3338 (1994).
\bibitem{g} M.Asakawa and C.M.Ko, {\it Nucl.Phys.} {\bf A572}, 732 (1994).
\bibitem{h} C.M.Ko and D.Seibert, {\it Phys.Rev.} {\bf C49}, 2198 
(1994).
\bibitem{i} U.G.Meissner, {\it Phys. Rep.} {\bf 161}, 213 (1988).
\bibitem{j} M.Bando, T.Kugo and K.Yamawaki, {\it Phys. Rep.}  {\bf 164}, 
217 (1988).
\bibitem{k} A.Bhattacharyya and S.Raha, {\it Phys.Lett.} {\bf B363}, 162 
(1995). 
\bibitem{l} B. A. Cole {\it et al.}, {\it Nucl.Phys.} {\bf A590}, 179c 
(1995).
\bibitem{m} M.Dey, V.L.Eletsky and B.L.Ioffe, {\it Phys.Lett.} {\bf B252}, 620 
(1990).
\end{thebibliography}
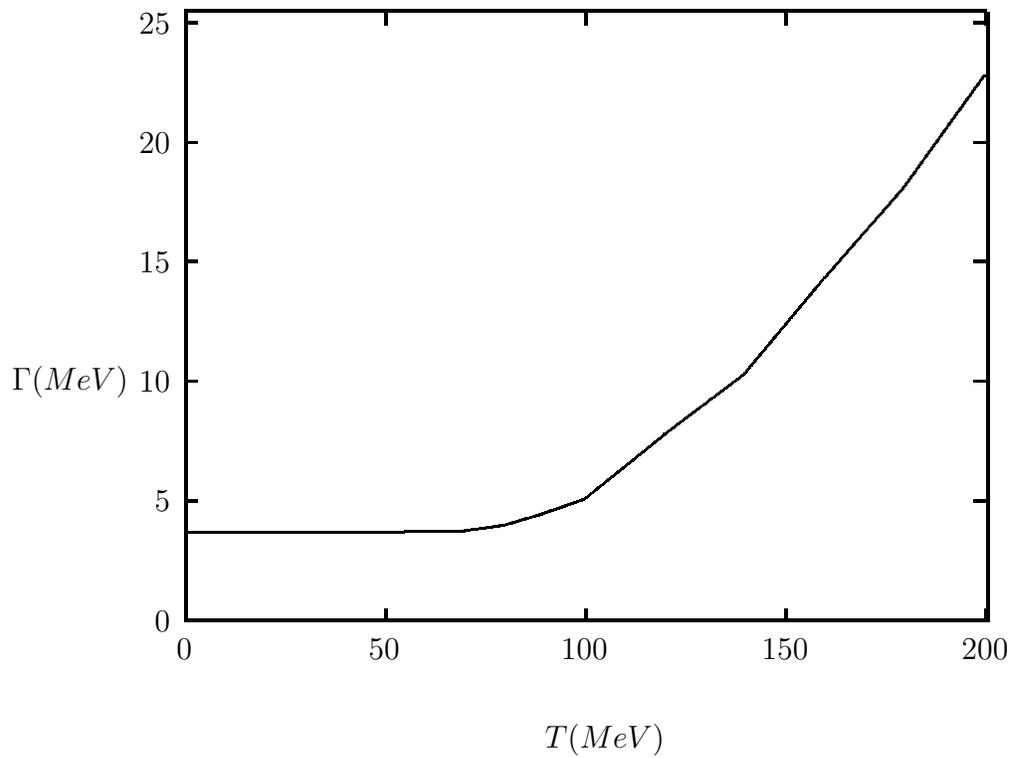
\end{document}